\documentclass[a4paper,11pt]{article}
\pdfoutput=1

\usepackage{jcappub}
\usepackage[T1]{fontenc}

\title{\boldmath Natural superheavy nuclei in astrophysical data}

\author[a,b,c,e]{A. Alexandrov,}
\author[d]{V. Alexeev,} 
\author[a]{A. Bagulya,}
\author[e]{A. Dashkina,} 
\author[a]{M. Chernyavsky,}
\author[a]{A. Gippius,} 
\author[a]{L. Goncharova,}
\author[a]{S. Gorbunov,}
\author[f]{V. Grachev,}
\author[d]{G. Kalinina,}
\author[a,e]{N. Konovalova,}
\author[a,e]{N. Okateva,}
\author[d]{T. Pavlova,}
\author[a,e,f,1]{N. Polukhina,\note{Corresponding author.}}
\author[h]{R. Rymzhanov,}
\author[a,e]{N. Starkov,}
\author[a,e]{T. N. Soe,}
\author[a,e]{T. Shchedrina,}
\author[a,g,h]{and A. Volkov}
\affiliation[a] {Lebedev Physical Institute, Russian Academy of
Sciences, 53 Leninsky Prosp., Moscow 119991, Russia}
\affiliation[b] {INFN sezione di Napoli, I-80126, Napoli, Italy}
\affiliation[c] {Universita' degli Studi di Napoli Federico II,
I-80126, Napoli, Italy}
\affiliation[d] {Vernadsky Institute of Geochemistry and Analytical
Chemistry, Russian Academy of Sciences, 19 Kosygin Str., Moscow 119991,
Russia}
\affiliation[e] {National University of Science and Technology
MISIS, 4 Leninsky Prosp., Moscow 119049, Russia}
\affiliation[f] {National Research Nuclear University MEPhI, 31
Kashirskoe Shosse, Moscow 115409, Russia}
\affiliation[g] {National Research Centre ``Kurchatov Institute'',
1 Kurchatov Sq., Moscow 123182, Russia}
\affiliation[h] {Flerov Laboratory of Nuclear Reactions, Joint
Institute for Nuclear Research, 6 Joliot-Curie Street, Dubna 141980,
Russia}

\emailAdd{andrey@na.infn.it}
\emailAdd{aval37@mail.ru}
\emailAdd{bagulyaav@lebedev.ru}
\emailAdd{asmailovab@gmail.com}
\emailAdd{mmchernyav@yandex.ru}
\emailAdd{gippiusaa@lebedev.ru}
\emailAdd{goncharova47@mail.ru}
\emailAdd{s.a.gorbunov@mail.ru}
\emailAdd{VMGrachev@mephi.ru}
\emailAdd{gakalin@mail.ru}
\emailAdd{ninakonovalova@yandex.ru}
\emailAdd{natalya\_okateva@mail.ru}
\emailAdd{pavlova4tat@mail.ru}
\emailAdd{Natalia.Polukhina@cern.ch}
\emailAdd{rymzhanov@jinr.ru}
\emailAdd{starkovni@lebedev.ru}
\emailAdd{dr.thannaingsoe.lpiras@gmail.com}
\emailAdd{tvshchedrina@gmail.com}
\emailAdd{a.e.volkov@list.ru}

\abstract{The paper presents the summary data of the authors' research within
the framework of the OLIMPIYA project (the Russian acronym of {\bf OLI}viny
iz {\bf M}eteoritov --- {\bf P}oisk tyazholykh {\bf I} sverkhtyazholykh
{\bf YA}der / Olivines from meteorites: Search for heavy and superheavy
nuclei) and results of track analysis for heavy cosmic ray nuclei
(\emph{Z} = 26--129) in olivine crystals from meteorites using an original
processing technique. A total of 21,743 tracks of nuclei heavier than
iron have been identified in meteoritic matter to date to form the
largest database within this charge range. The database includes three
tracks of superheavy nuclei with the lifetimes of about a few decades,
which can be considered as direct experimental evidence for the existence
of natural superheavy nuclei from the ``island of stability''. Comprehensive
comparative analysis of data from two meteorites with different cosmic
ray exposure ages, Marjalahti (from 178 to 205 Myr) and Eagle Station
(from 35 to 71 Myr), is presented for the first time. The results
are discussed within the existing concepts of nuclei formation in
astrophysical processes.}

\keywords{superheavy nuclei, galactic cosmic ray, olivine, image recognition}

\begin{document}
\maketitle
\flushbottom

\section{Introduction}
\label{sec:intro}

The paper presents a summary of the authors' work on the use of meteoritic
olivines as effective detectors of extraterrestrial nuclei. Emphasis
is made on the search for heavy and superheavy nuclei in galactic
cosmic rays and the analysis of possible ways of their formation and
propagation in the Universe. The work is directly related to the fundamental
problem of the limits of the Periodic Table of Elements, the 150th
anniversary of which is celebrated in 2019 by the UNESCO decision.

Although the standard electrodynamics allows for nuclei with atomic
numbers greater than 170~\cite{Nls69}, the only massive natural stable
chemical element on Earth is uranium with a nucleus charge of 92;
all nuclei heavier than that have been obtained artificially. Instability
of nuclei heavier than uranium results from a faster increase of Coulomb
repulsion of protons in comparison with attraction caused by nuclear
forces when the number of nucleons in a nucleus increases. The nuclear
shell model~\cite{Goe64} predicts ``islands of stability'' for superheavy
nuclei. According to theory, the arrangement of nucleons into complete
shells within the atomic nucleus creates states of the largest binding
energy, which for the so called ``magic'' numbers of neutrons and
protons (2, 8, 20, 50, 82, 126 --- calcium, tin, lead, etc.) significantly
increase the height of the nuclear fission barrier~\cite{Nls69,Mos69,Ack17}.
The half-lives of the stable isotopes of these elements can be from
several minutes to several years (up to thousands of years, according
to over-optimistic estimates~\cite{Ogn17}). Attempts to produce new superheavy
elements and to synthesize elements with \emph{Z} $\ge$ 100 continue on
accelerators~\cite{Uty16,Sea90,Iup}.

Synthesis of superheavy elements poses certain experimental problems.
All elements with \emph{Z} $\ge$ 100 have been produced on high-power
accelerators in heavy ion-induced fusion reactions. The probability
of these reactions to occur and their products to survive fission
is extremely low due to the necessity to create extremely high energy
densities and high neutron fluxes, so accelerator experts try to make
a proper choice of target projectile and projectile's energy combinations.
The time required to synthesize one atom of elements 119 or 120 is
estimated to be hundreds of days~\cite{Hab19}. The challenge is aggravated
by the fact that laboratory-produced superheavy elements are represented
by neutron-poor isotopes, which are generally unstable~\cite{Ogn15}. The
resulting dominance of Coulomb repulsion leads to a spatial heterogeneity
and non-sphericity of the nuclei, thus stimulating their decay.

If theory is right, and the ``island of stability'' does exist,
it should be possible to detect these nuclei in nature. Astrophysical
studies are recognized as a promising way to search for these nuclei~\cite{Ter15,Dzu17}. As the fraction of heavy element nuclei in cosmic rays
is very small and does not exceed 0.5\%, their flux in near-Earth
space is about a mere 1--2 nuclei/m$^2$ per year. This
results in an altogether small number of nuclei from the region of
transuranium elements with single events related to nuclei with \emph{Z}
> 92 registered in direct balloon~\cite{Fow70,Pri71,Cec10} and satellite~\cite{Shi78,Fow87,Bin89,Don12} experiments with cosmic rays. This limitation can to a
significant extent be overcome by using natural track detectors, which
are meteorites and lunar samples~\cite{Rao70,Bha73}. Estimates by Prof.
G.N. Flerov show~\cite{Fle77} that the study of one gram of mineral from
a meteorite with a cosmic ray exposure (CRE) age of $\sim${}10$^{8}$ years is equivalent to an experiment for the direct irradiation of
a ton of nuclear emulsion for one year on the Earth orbit.

Study of olivine crystals from the Marjalahti and Eagle Station pallasites
for direct registration of tracks of heavy nuclei was initiated in
the 1970s at the Flerov Laboratory of Nuclear Reactions, Joint Institute
for Nuclear Research (Dubna)~\cite{Fle81}. These experiments resulted
in fixation of several super-long tracks of nuclei heavier than uranium~\cite{Per03}. In 2005, two fragments of these meteorites were transferred
to the Lebedev Physical Institute of the Russian Academy of Sciences
to carry out further studies using new techniques and apparatus~\cite{Gin05}.
In subsequent years, this research, being carried out together with
colleagues from the Vernadsky Institute of Geochemistry and Analytical
Chemistry, Russian Academy of Sciences, provided for accumulation
of a large database of heavy and superheavy GCR nucleus tracks in
meteoritic olivine~\cite{Ale16,Ale17}. Analysis of these data confirmed
the presence of tracks of transuranium and transfermium elements in
meteoritic matter. Our paper presents the summary results of the research
within the framework of the OLIMPIYA project to search for and identify
heavy and superheavy nuclei of galactic cosmic rays (GCR) and confirm
experimentally the existence of transuranium and transfermium elements
in nature.

Two results presented in the paper motivated its writing. First, we
obtained new results of track processing, which considerably increase
the database statistics. Second, we compared data from two meteorites
of different radiation ages, Marjalahti and Eagle Station, to enable
an experimental database to be set up to test models of the formation
of heavy and superheavy elements in various sources and to identify
the main mechanisms of their appearance as the GCR fraction.

\section{Olivine crystals as natural track detectors}
\label{sec:olivine}

The use of stony-iron meteorites as natural detectors for the search
for superheavy nuclei in nature, namely in galactic cosmic rays, seems
promising for a number of reasons. First, the exposure times of meteorites
are millions of years~\cite{Her15}, which increases the probability of
fixing superheavy nuclei, the fraction of which in cosmic rays is
negligible. Second, since meteorites travel far beyond the Solar System,
they can capture superheavy nuclei in relative proximity to the source
of their formation, in contrast to near-Earth detectors. And, in addition
to stable nuclei, meteorites are capable of fixing unstable isotopes
of superheavy nuclei with lifetimes longer than their times of flight
from the source to meteorite.

The presented research makes use of natural track detectors, which
are translucent crystals of magnesium-iron silicate olivine (Mg$_{0.8}$Fe$_{0.2}$)$_{2}$SiO$_{4}$ extracted from Marjalahti and Eagle Station pallasites and prepared for microscopic measurements. Contrary to other types of track detectors (photoemulsion, plastic), olivine has no background tracks from nuclei lighter than
iron, which is due to a rather high threshold of specific energy losses
of the charged projectile required to form its chemically etched track
in olivine (the threshold value of ionization losses of energy for
olivine is about 18 MeV/(mg$\cdot$cm$^{2}$)~\cite{Hor67}).

Research into tracks of GCR nuclei in olivines from pallasites is
based on an original method of controlled etching~\cite{Ale16,Bag13} exploiting
the fact that areas of a crystal damaged by a passing charged particle
have a higher chemical activity compared to intact areas. The dissolution
of material by a multicomponent chemical etchant proceeds at a much
higher rate in this area to enable the ``development'' of nuclear
tracks making them visible in translucent olivine under an optical
microscope and available for analysis. Identification of cosmic ray
nuclei in olivine crystals is based on the study of the dynamic and
geometric characteristics (etching rate \emph{V} and residual path
length \emph{L}) of chemically etched segments of the nuclei deceleration
trajectories before the nuclei are completely stopped in olivine crystals.
These characteristics depend on the extent of crystal lattice damage
and on ionization caused by the passage of a swift heavy ion and,
therefore, on the magnitude of its charge.

The polished surface of olivine crystals is subjected to multistep
chemical etching and processing; the parameters of tracks visible
at a given treatment step are measured. After the measurements, a
50--100 $\mu$m crystal layer (accurate to $\mu$m) is removed, and the procedure
is repeated. The new detected tracks are being linked with those in
the previous layer. The complete track lengths are summed up with
account for the trajectory slope to the crystal surface and the thickness
of the removed layer. Such successive operations of polishing/etching/search/track
registration enable the fixation of a) the spatial orientation of
tracks in a crystal, b) the nucleus stopping point or, which is the
same, residual range \emph{L} of a projectile and c) the etching rate
\emph{V} at various segments of the nucleus trajectory.

Search for and analysis of tracks is performed on the PAVICOM automated
measuring setup~\cite{Alk04}. Figure~\ref{fig:a} presents micrographs (with 60$^\times$
magnification) of typical etched nucleus tracks in the developed olivine
crystals. Reflecting the dependence of the energy loss on the energy
of a projectile, the tracks are formed of two parts: cylindrical,
at the end of the trajectory; and needle-like, at the trajectory segment
corresponding to larger projectile energies. Figure~\ref{fig:a}(a) clearly shows
the boundary separating the track into two segments, the channel diameters
of which differ severalfold; in figure~\ref{fig:a}(b), this transition is less
pronounced.

\begin{figure}[h]
\begin{minipage}[h]{0.49\linewidth}
\center{\includegraphics[width=1.\linewidth]{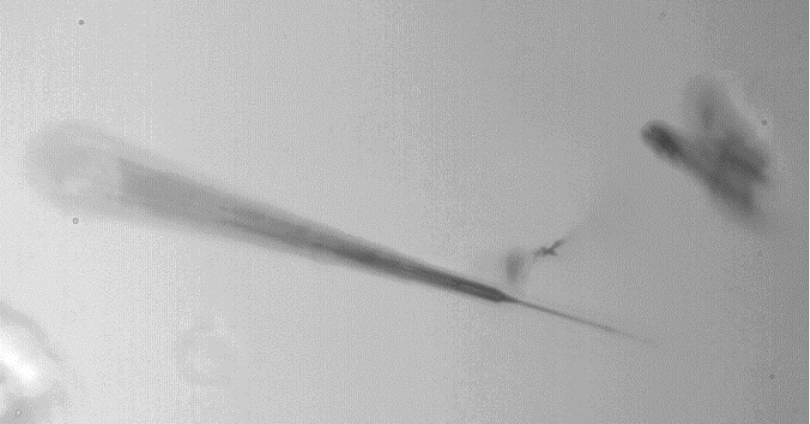} \\ a)}
\end{minipage}
\hfill
\begin{minipage}[h]{0.49\linewidth}
\center{\includegraphics[width=0.9\linewidth]{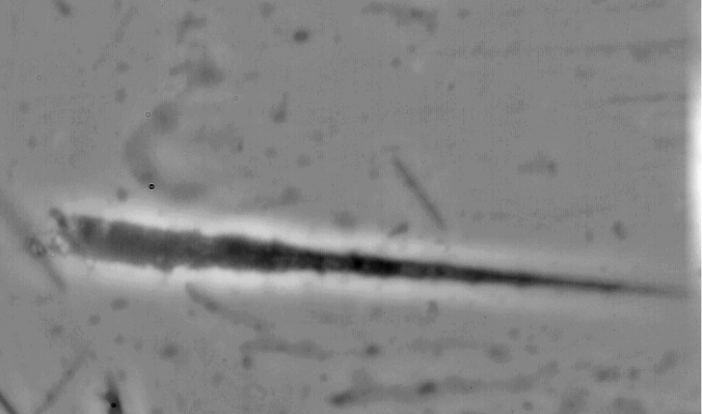} \\ b)}
\end{minipage}
\caption{\label{fig:a} Micrographs of GCR superheavy-element nuclear tracks registered in olivine crystals from pallasites. Field of view 150$\times$ 80
$\mu$m (a) and 100$\times$ 55 $\mu$m (b).}
\label{fig:a}
\end{figure}

Irradiation of olivine crystals with different swift heavy ions on
accelerators of IMP (Lanzhou, China) and GSI (Darmstadt, Germany)~\cite{Bag13}, as well as comparison with the results of simulations based
on an original numerical model of track formation and etching in olivine~\cite{Gor15}, allowed for charge calibration of the data obtained in the
OLIMPIYA experiment. These studies and subsequent analysis gave us
an opportunity to study in detail the mechanisms involved in the formation
of detected tracks as well as to obtain the dependence of the etching
rate on the residual range and nucleus charge \emph{V}(\emph{L},\emph{Z}),
which provided for the determination of charges of the registered
nuclei to an accuracy of $\pm$ 1 to $\pm$ 2 units for \emph{Z}~<~92.
Within the range of 67 < \emph{Z} < 92, the
dependence was fitted by a five-parameter function allowing for its
extrapolation to larger \emph{Z} values~\cite{Ale16}.

A large database of the processed tracks of GCR nuclei in olivines
from Marjalahti and Eagle Station pallasite meteorites has been accumulated
over about a decade of the experiment. Our results are widely used
in discussions on the origin of galactic cosmic rays and on the hypothesis
of the existence of the ``island of stability'' nuclei~\cite{Hab19,Dzu17,Kom17,Tho19,Ban19,Giu19}.

\section{Results and discussion}
\label{sec:res}

According to the current views on the origin of heavy and superheavy
nuclei in the Universe, elements heavier than iron can be formed by
two scenarios --- slow or rapid neutron-capture processes (\emph{s}-processes
and \emph{r}-processes, respectively)~\cite{Bur57,Gor17}. As the result
of both these processes, the nucleus rich in neutrons experiences
a $\beta$-decay that converts neutrons into protons and increases the nucleus
charge. In an \emph{s}-process, production of nuclei is based on the
gradual capture of neutrons with subsequent $\beta$-decays, i.e. the process
moves along the chain of re-emerging short-lived proton-enriched nuclei.
The process is ``slow'' because it needs sufficient time for the
realization of the radioactive decay before the next neutron will
be captured by a nucleus. The \emph{s}-process is realised at ``low''
neutron fluxes/densities (10$^{5}$ to 10$^{11}$ cm$^{2}$s$^{-1}$) and is not able to produce heavy elements adjacent to uranium.

At a neutron density of 10$^{10}$ n/m$^{3}$, the transformation of Fe to Pb takes place. The \emph{s}-process responsible for the formation of elements up to \emph{Z} = 83 can occur, in particular, in the Red Giants. Nuclei in the range of \emph{Z} = 84--89 do not have stable isotopes, that is why it is impossible to ``fill'' this segment of GCR charge spectrum with \emph{s}-process products. On
the other hand, nuclei with \emph{Z} = 90 (thorium) and \emph{Z} =
92 (uranium isotopes $^{235}$U and $^{238}$U)
do exist in nature. This can be explained by the assumption that they
are formed by the consecutive capture of several neutrons by unstable
nuclei in the \emph{r}-process.

The \emph{r}-process is a successive neutron capture occurring faster
than the $\beta$-decay does. The heaviest nuclei are formed in a ``strong''
(or ``violent'') \emph{r}-process at neutron densities higher than
10$^{19}$~cm$^{–3}$ (and reaching 10$^{24}$ or even 10$^{27}$ cm$^{-3}$), when the
nucleus of iron captures 100 neutrons and more. According to different
theoretical scenarios, conditions for superheavy nuclei formation
can be realized in supernovae (SNs)~\cite{Win12} or in neutron stars or
else neutron star-black hole mergers (the latter two both indicated
as NSMs)~\cite{Bau14,Thi17}. NSMs are considered to be the most powerful
sources of \emph{r}-process matter, ejecting on average from 100 to
1000 times more \emph{r}-process material than SNs do. Therefore,
although the NSMs (manifesting themselves as macronovae or kilonovae)
occur significantly more rarely than core-collapse supernovae (the
frequency of SN collapses in the Milky Way Galaxy is $\sim$1/30
y$^{-1}$, i.e., several orders of magnitude higher than
the frequency of NSM events, which is 10$^{-5}$ y$^{-1}$), they could potentially be the dominant mode of producing heavy elements~\cite{Kom17,Gor13}.

Synthesis of superheavy elements in kilonovae is confirmed by recent
data on the electromagnetic spectrum of the event GW170817 interpreted
as a signal from an NSM, in which two components of the emission are
distinguished; one of them consists mainly of light (atomic mass number
less than 140), the other of heavy elements of the \emph{r}-process
(atomic mass number more than 140)~\cite{Kas17}. These observations became
available through the discovery of gravitational waves from inspiralling
neutron stars in the LIGO-Virgo experiments ~\cite{Abb17}, confirming and
significantly improving the sky localization of this event.

Because of a large difference between the exposure times of meteorites
and artificial orbital satellites, they must register nuclei synthesized
in different sources. The satellites are exposed to current fluxes
of nuclei in near-Earth space and register mainly fragments of nuclei
formed in neutron matter fusion in relatively frequent SN explosions.
In close vicinity to the Solar System (1.5--3 kpc), i.e. at distances
from which the formed nuclei keeping information about details of
their birth come, only several NSM events occur in 100 million years,
which means that the exposure times of orbital satellites are too
small to fix nuclei from NSMs. Meteorites can fix bursts of fluxes
of superheavy nuclei born in NSMs in the \emph{r}-process of neutron
matter fusion, in relative vicinity to which they occurred when exposed
to radiation formed. In this way, meteorites hit by superheavy element
ejecta from NSMs become witness of cosmic events that had occurred
millions of years ago hundreds of light years away from the Earth.

This hypothesis is substantiated in~\cite{Kom17}, based in particular
on the data obtained by the OLIMPIYA, Ariel-6~\cite{Fow87}, HEAO-3~\cite{Bin89},
UHCRE~\cite{Don12} and SuperTIGER~\cite{Mur16} experiments. The authors of~\cite{Kom17} performed model calculations for the evolution of the energy
spectrum of superheavy nuclei formed in an NSM \emph{r}-process for
various models of acceleration of particles in space. The analysis
given in this work has shown that the OLIMPIYA data are in good agreement
with predictions of the NSM model.

\begin{figure}[tbp]
\centering 
\includegraphics[width=0.6\linewidth]{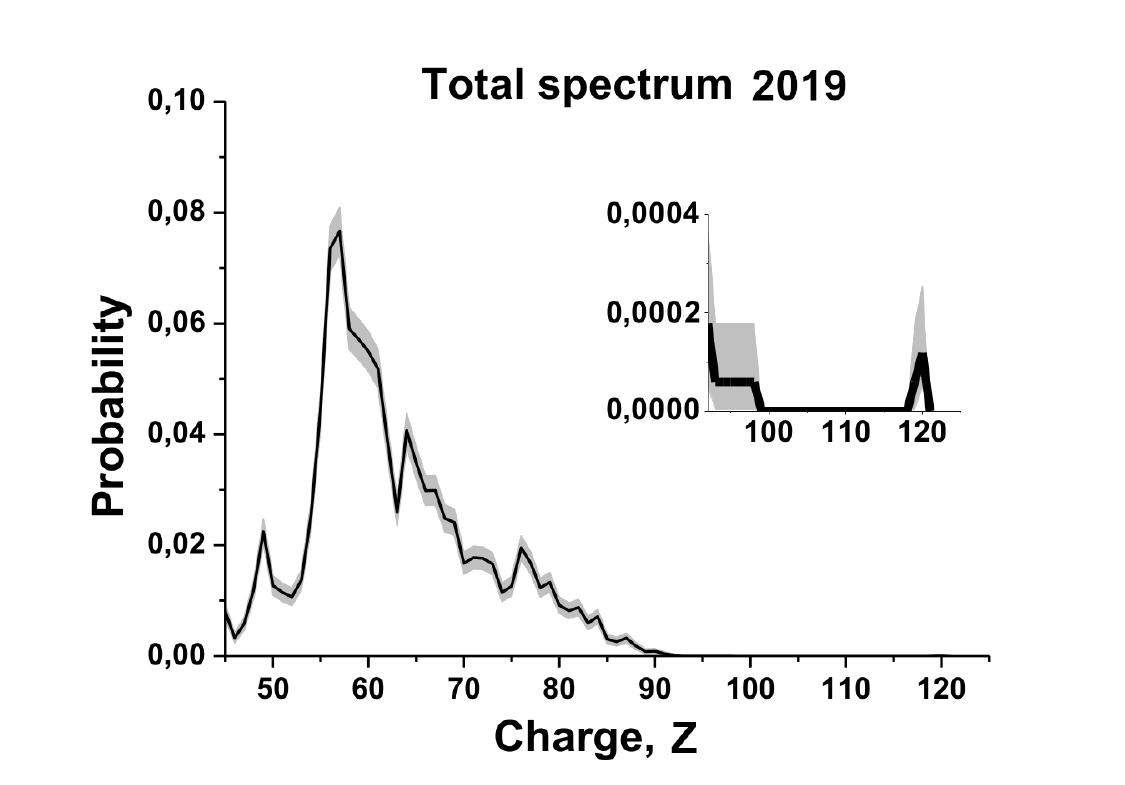}
\caption{\label{fig:b} Renewed relative distributions with the error bars (95\% reliability) for the nuclei detected in both meteorites in the OLIMPIYA experiment.}
\end{figure}

\begin{figure}[h]
\centering 
\includegraphics[width=0.6\linewidth]{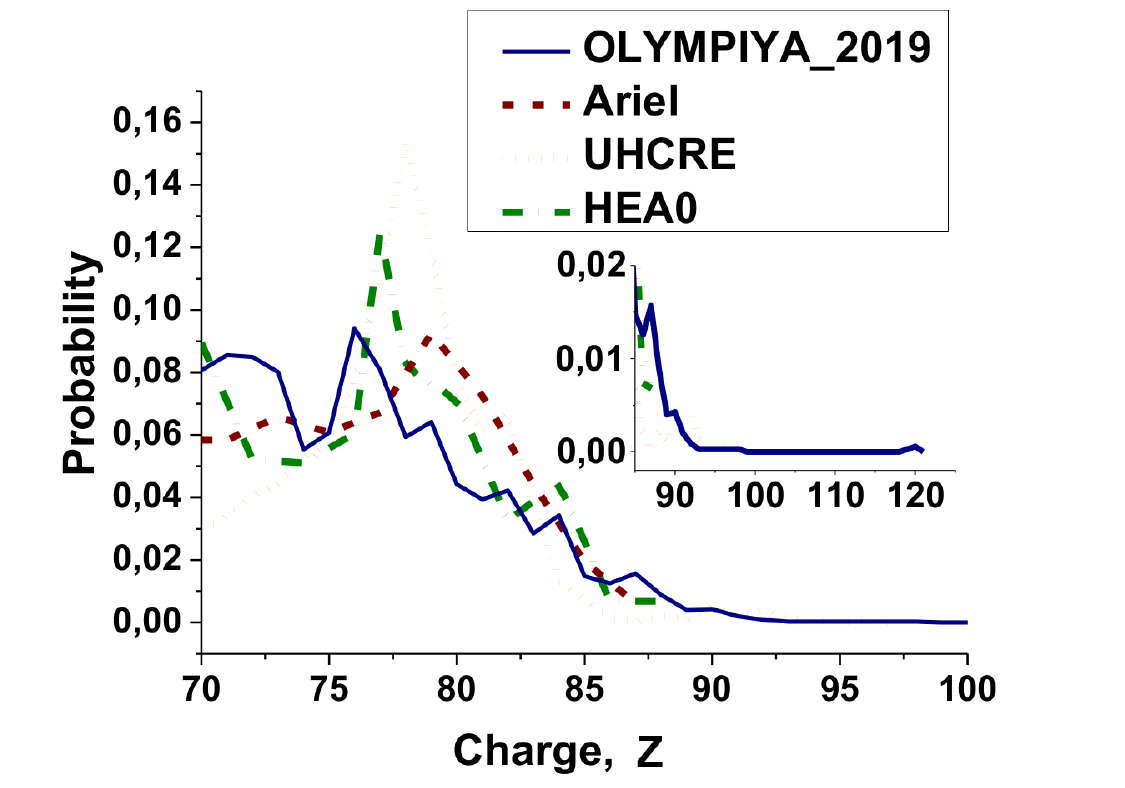}
\caption{\label{fig:c} Relative abundance of GCR nuclei in our experiment OLIMPIYA
compared with Ariel-6~\cite{Fow87}, HEAO-3~\cite{Bin89} and UHCRE~\cite{Don12}
experiments.}
\end{figure}

\begin{figure}[tbp]
\centering 
\includegraphics[width=0.6\linewidth]{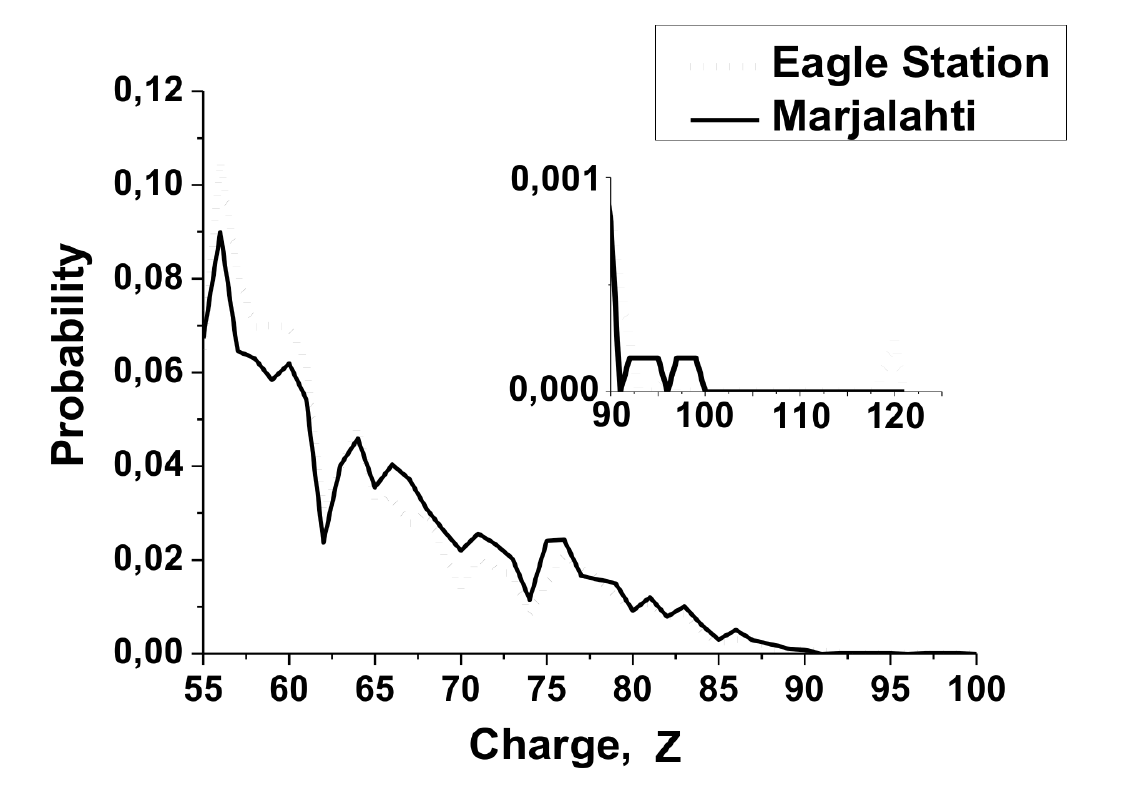}
\caption{\label{fig:d} Difference between the charge spectra of heavy nuclei detected
in the Marjalahti (exposure time from 178 to 205 Myr) and Eagle Station
(exposure time from 35 to 71 Myr) meteorites.}
\end{figure}

Since our last comprehensive review~\cite{Ale16}, we have been able to
significantly improve the reliability of the experimental results,
almost doubling the statistics of the measured tracks. The statistics
accumulated during this period amounted to 21,743 processed tracks
of heavy and superheavy nuclei with charges from 26 (the threshold
value) to, at least, 119. Figure~\ref{fig:b} presents the renewed data of the
relative distributions of the nuclei recorded in both meteorites with
95\% reliability error bars. (In the next figures all the data are
given without the error bars in order to facilitate the required comparisons.)
The data in figures~\ref{fig:b},~\ref{fig:c},~\ref{fig:d} are normalized so that the sum of all probabilities
over all integer charges gives unity. As in~\cite{Ale16}, all our results
are normalized to the abundance of iron nuclei \emph{A}($_{26}$Fe)
= 10$^{6}$.

Table~\ref{tab:a} and figure~\ref{fig:c} demonstrate the difference between our updated
data and the results of the satellite experiments~\cite{Fow87,Bin89,Don12}. Comparison
of the nuclei abundance ratios demonstrates that the fractions of
superheavy nuclei registered in the meteorites significantly (up to
100\%) exceed those registered by satellite instruments on the Earth
orbit, i.e. indicates that these nuclei were apparently born in \emph{r}-processes
that took place in very rare NSM events. Table~\ref{tab:a} and figure~\ref{fig:d} also
show the difference between the relative abundances of nuclei registered
in Marjalahti and Eagle Station.

\begin{table}[tbp]
\centering
\begin{tabular}{lccccc}
\hline
  & HEAO-3 & Ariel-6 & UHCRE & OLIMPIYA & OLIMPIYA \\
  & (Binns 1985, & (Fowler & (Donnelly & (Marjalahti) & (Eagle \\
  & 1989) & 1984, & 2012) &  & Station) \\
  &  & 1987) & & & \\
\hline
Actinides/Pt  & 0.0241$_{-0.010}^{+0.022}$  & ---  & 0.025$\pm$ 0.005  & 0.05$\pm$ 0.008  & 0.038$\pm$ 0.008 \\
Actinides/Subactinides  & 0.0186$_{-0.010}^{+0.018}$  & 0.04  & 0.020$\pm$ 0.004  & 0.036$\pm$ 0.006  & 0.028$\pm$ 0.006\\
Pb/Pt  & 0.30$\pm$ 0.081  & 0.40$\pm$ 0.10  & 0.25$\pm$ 0.04  & 0.23$\pm$ 0.026  & 0.24$\pm$ 0.03\\
HS$^{{*}}$/PbPt  & 0.16$\pm$ 0.06  & 0.27$\pm$ 0.07  & 0.19$\pm$ 0.03  & 0.64$\pm$ 0.03  & 0.52$\pm$ 0.035\\
LS$^{{*}{*}}$/PbPt  & 1.1$\pm$ 0.2  & 1.45$\pm$ 0.25  & ---  & 1.98$\pm$ 0.065  & 2.1$\pm$ 0.13\\
\hline
\end{tabular}
\footnotesize{$^{{*}}$ Heavy Secondary (70 $\le$ \emph{Z} $\le$ 73).\\
$^{{*}{*}}$ Light Secondary (62 $\le$ \emph{Z} $\le$ 69).}
\caption{\label{tab:a} Abundance ratios based on the satellite experiments~\cite{Fow87,Bin89,Don12} and the OLIMPIYA updated data.}
\end{table}

%\begin{figure}[tbp]

The number of nuclei that passed through meteoritic matter depends
on the CRE age of a meteorite. The estimated exposure times of the
meteorites used in our research range from 35 to 71 Myr for Eagle
Station and from 178 to 205 Myr for Marjalahti~\cite{Bag15}. These times
are much larger than 17 Myr which, according to estimations made in~\cite{Kom17}, are required to fix superheavy nuclei by a meteorite. Due
to a considerable difference between their CRE ages, these two meteorites
have witnessed, with high probability, a different number of NSM events.
Citing~\cite{Kom17}, ``if the time averaged fluxes measured by meteorites
with different ages show different values, it will be a smoking gun
of the NSM contribution of UHCRs'' (ultra-heavy cosmic rays). Now
we can conclude that the difference between the abundance ratios of
superheavy nuclei registered in Eagle Station and Marjalahti (see
Table~\ref{tab:a}) looks like a direct indication of this smoking gun of proof.

As for the data of two meteorites, Figure 4 shows that the probability
of registering nuclei with 50 < \emph{Z} < 60 for Eagle Station is higher than for Marjalahti; for \emph{Z} > 65, it is the other way round. Thus, the average slope of the charge dependence decreases with exposure time and could be explained by the low abundance of the heaviest components in cosmic rays. This
fact is also manifested in the abundance of transuranium nuclei (see
Table 1); it is much higher in Marjalahti than in Eagle Station, which
could be expected due to a longer exposure time of the former meteorite.
The observed difference in the abundance of heavy nuclei can also
be explained by different distances from the meteorites to the NSM
during their irradiation.

The experimentally recorded difference between the relative abundances
of nuclei registered in meteorites with different exposure times,
as well as the difference between the meteorite data and the results
of the satellite experiments, can be used to test models of nuclei
propagation to the Solar System, including fragmentation effects of
heavy elements generated by NSMs.

The most significant result of the OLIMPIYA experiment is the detection
and identification of three transfermium nuclei with charges estimated
as 119$_{-6}^{+10}$,
which can be considered as the first direct evidence of the existence
of naturally occurring stable superheavy nuclei. In 2013, the OLIMPIYA
group reported the detection of three tracks of particles with atomic
numbers in the interval from 105 up to 130~\cite{Bag13}. The track lengths
corresponding to the registered superheavy nuclei exceeded 500 $\mu$m,
and the etching rates \emph{V} were greater than 35 $\mu$m/h~\cite{Ale16}.
Taking into account that the experimentally measured maximum rate
of track etching for uranium nuclei in olivines before their stoppage
is 26$\pm$1 $\mu$m/h, it becomes clear that the charges of these nuclei
significantly exceed the charge of uranium. The charge assessment
for these nuclei was based on the dependence of the etching rate near
the stoppage point on the charge value (see Section~\ref{sec:olivine}). The dependence of the etching rate on the charge was approximated by a straight line
along the available five experimental points up to the value of \emph{Z}
= 92. Further extrapolation of this straight line up to the etching
rate value of 35 $\mu$m/h and the regression analysis performed made possible
the assessment of charge of these nuclei with 95\% probability.

In the context of the above, it can be argued with a high degree of
certainty that the data obtained in the OLIMPIYA experiment supply
arguments supporting the existence of the theoretically predicted
``island of stability'' of long-lived transfermium nuclei in nature.
The three ultraheavy nuclei detected can be considered as experimental
confirmation of the existence of transfermium elements in nature and
represent a result of high significance.

Of great interest are the chemical properties of the elements closest
to $^{118}$Og with larger nucleus charges, not yet obtained
on accelerators. The peculiarities of the electronic structure of
atoms with superheavy nuclei are determined by relativistic effects
and may differ sharply from the ``usual'' electronic properties
of atoms with less heavy nuclei. Since $^{118}$Og is an
inert gas (or liquid) and completes a period of the Periodic System
of Chemical Elements, elements 119 and 120 should belong to the 8s-block
and have the properties of alkaline and alkaline-earth ``metals'',
respectively. But substances composed of superheavy atoms may exhibit
still unknown and unusual chemical and physical properties. Prediction
and registration of these properties pose a fundamental problem.

\section{Conclusion}
\label{sec:concl}

The presented material demonstrates the efficiency of the meteorite
method for studying the heavy component of galactic cosmic rays. To
date, 21,743 tracks of ions heavier than iron have been detected and
identified in the OLIMPIYA experiment, providing statistically significant
data on the relative abundance of heavy and superheavy nuclei in galactic
cosmic rays. These data are compared with the results obtained in
satellite experiments, as well as in two meteorites with different
ages of exposure to cosmic rays. The results have been obtained and
are discussed within the framework of the existing concepts of the
formation of heavy and superheavy nuclei in astrophysical processes
in the Galaxy. The difference between nuclei abundances in two meteorites
is a strong evidence of the superheavy nuclei formation in mergers
of neutron stars (or neutron star-black hole mergers) and support
the hypothesis that the condition for their synthesis is a huge free
neutron density granting a rapid capture of neutrons and suppressing
$\beta$-decay.

One of the most important results of the experiment, which is the
registration of three tracks of nuclei with charges in the range of
113 < \emph{Z} < 129, has been confirmed.

\acknowledgments

We express our deep gratitude to Professor Yu. Ts. Oganesyan for useful
discussion and support, which included the provision of meteorite
samples for the research.

This work was carried out using computing resources of the Federal
Collective Usage Center ``Complex for Simulation and Data Processing
for Mega-science Facilities'' at the NRC Kurchatov Institute (ministry
subvention under agreement RFMEFI62117X0016), \url{http://ckp.nrcki.ru/}.

\end{document}